\documentclass[twocolumn,aps,prl,amssymb,amsmath,amsfonts,superscriptaddress,showpacs,prl, reprint, superscriptaddress, showpacs, bibliography]{revtex4-1}
\usepackage{graphicx}
\usepackage[usenames,dvipsnames]{xcolor}
\usepackage{bm}
\usepackage{wasysym}
\usepackage{color}
\def\be{\begin{equation}}
\def\ee{\end{equation}}
\def\bc{\begin{center}}
\def\ec{\end{center}}
\newcommand{\lcvo}{LiCuVO$_4$}

\newcommand{\bea}{\begin{eqnarray}}
\newcommand{\eea}{\end{eqnarray}}
\usepackage[colorlinks,bookmarks=false,citecolor=red,linkcolor=red,urlcolor=NavyBlue]{hyperref}
\begin{document}
\title{Nuclear magnetic resonance signature of the spin-nematic phase \\ in \lcvo~at high magnetic fields}
\author{A. Orlova}
\email{anna.orlova@lncmi.cnrs.fr}
\affiliation{Laboratoire National des Champs Magn\'etiques Intenses, LNCMI-CNRS, UGA, UPS, INSA, EMFL, 31400 Toulouse and 38042 Grenoble, France}
\author{E. L. Green}
\email{e.green@hzdr.de}
\affiliation{Dresden High Magnetic Field Laboratory (HLD-EMFL), Helmholtz-Zentrum Dresden-Rossendorf, 01314 Dresden, Germany}
\author{J. M. Law}
\altaffiliation[Present address: ]{CryoVac GmbH \& Co KG, 53842 Troisdorf, Germany}
\affiliation{Dresden High Magnetic Field Laboratory (HLD-EMFL), Helmholtz-Zentrum Dresden-Rossendorf, 01314 Dresden, Germany}
\author{D. I. Gorbunov}
\affiliation{Dresden High Magnetic Field Laboratory (HLD-EMFL), Helmholtz-Zentrum Dresden-Rossendorf, 01314 Dresden, Germany}
\author{G. Chanda}
\affiliation{Dresden High Magnetic Field Laboratory (HLD-EMFL), Helmholtz-Zentrum Dresden-Rossendorf, 01314 Dresden, Germany}
\author{S.~Kr\"amer}
\affiliation{Laboratoire National des Champs Magn\'etiques Intenses, LNCMI-CNRS, UGA, UPS, INSA, EMFL, 31400 Toulouse and 38042 Grenoble, France}
\author{\mbox{M. Horvati\'c}}
\affiliation{Laboratoire National des Champs Magn\'etiques Intenses, LNCMI-CNRS, UGA, UPS, INSA, EMFL, 31400 Toulouse and 38042 Grenoble, France}
\author{R. K. Kremer}
\affiliation{Max-Planck Institute for Solid State Research, Heisenbergstra{\ss}e 1, D-70569 Stuttgart, Germany}
\author{J. Wosnitza}
\affiliation{Dresden High Magnetic Field Laboratory (HLD-EMFL), Helmholtz-Zentrum Dresden-Rossendorf, 01314 Dresden, Germany}
\affiliation{Institut f\"ur Festk\"orperphysik, TU Dresden, 01062 Dresden, Germany}
\author{G. L. J. A. Rikken}
\affiliation{Laboratoire National des Champs Magn\'etiques Intenses, LNCMI-CNRS, UGA, UPS, INSA, EMFL, 31400 Toulouse and 38042 Grenoble, France}
\date{\today}

\begin{abstract}
We report a $^{51}$V  nuclear magnetic resonance investigation of the frustrated spin-1/2 chain compound \lcvo, performed in pulsed magnetic fields and focused on high-field phases up to 55~T. For the crystal orientations $H \parallel c$ and $H \parallel b$ we find a narrow field region just below the magnetic saturation where the local magnetization remains uniform and homogeneous, while its value is field dependent. This behavior is the first microscopic signature of the spin-nematic state, breaking spin-rotation symmetry without generating any transverse dipolar order, and is consistent with theoretical predictions for the \lcvo~compound.
\end{abstract}

\pacs{75.10.Kt, 75.30.Kz, 76.60.-k}

\maketitle

%

The search for new states of quantum matter is one of the most active research fields in condensed-matter physics. In this respect frustrated magnetic systems attract a lot of interest as they accommodate various unconventional quantum states, having no direct classical analogues, ordered and disordered, induced by the competing interactions \cite{Lacroix_2011}. One particularly interesting state is the spin-nematic phase, in which the quantum magnet behaves like a liquid crystal. Taking an external magnetic field $H$ as the reference direction, a spin-nematic phase is defined as a state without any transverse dipolar (i.e., vector-type) order, \mbox{$(-1)^{i}\langle S^{+}_{i} + \rm{H.c.}\rangle$ = 0}, but possessing instead a transverse \textit{quadrupolar} (tensor-type) order, \mbox{$(-1)^{i}\langle S^{+}_{i}S^{+}_{i+1} + \rm{H.c.}\rangle~\neq$ 0}. The quadrupolar order parameter develops on the bonds between neighboring  spins and can be described  as a condensate of two-magnon pairs. It breaks the spin-rotational symmetry about the magnetic field, but only partially as \mbox{$\pi$- rotations} transform the order parameter into itself. The also broken translational symmetry of the order parameter is invisible in the dipolar channel. There is also an analogy between the spin-nematic phase and the superconducting state: the nematic phase can be considered as a ``bosonic'' superconductor formed as a result of \textit{two-magnon} condensation \cite{Lacroix_2011, Starykh_2015}.

The concept of a spin-nematic state was developed by Andreev and \mbox{Grishchuk} more than 30 years ago \cite{Andreev_1984}, which incited intense search for a realization in real materials. However, a definite experimental proof for the existence of such a phase has not been provided yet. Several magnetic insulators have been proposed as possible candidates, including the two-dimensional magnet NiGa$_2$S$_4$ (spin-1 system) \cite{Nakatsuji_2005,Takeya_2008,Nakatsuji_2010} and thin films of $^{3}$He \cite{Roger_1983,Momoi_2006,Momoi_2012}.

In the past 10 years a large number of theoretical studies have supported the formation of the spin-nematic phase in frustrated zig-zag 1D (chain) systems \cite{Shannon_2006,Ueda_2013,Sato_2013,Starykh_2014,Onishi_2015}. Amongst these, orthorhombic \lcvo~is one of the most promising candidates \cite{Svistov_2011,Zhitomirsky_2010}. It consists of \mbox{spin-1/2} Cu$^{2+}$ chains along the orthorhombic $b$ axis with a dominant nearest-neighbor ferromagnetic interaction $J_{1}=-1.6$~meV, a frustrated next-nearest-neighbor antiferromagnetic interaction $J_{2}=3.8$~meV, and an interchain coupling $J=-0.4$~meV \cite{Prokofiev_2004, Enderle_2005}. At zero magnetic field an incommensurate planar spiral structure is realized below $T_{N}$ = 2.3~K, having the moments lying in the $ab$ plane \cite{Heidrich_Meisner_2009, Enderle_2010}. Above 10 T, an incommensurate, \textit{collinear} spin-density wave (SDW) phase is stabilized \cite{B_ttgen_2007,B_ttgen_2010,B_ttgen_2012,Nawa_2013}. Neutron-diffraction experiments show that it consists of bound magnon pairs with $S_{z}=2$ that form a periodic structure \cite{Masuda_2011,Mourigal_2012}. The temperature dependence of the NMR spin-relaxation rate in that phase reveals an energy gap developing in the transverse spin-excitation spectrum below the magnetic ordering temperature $T_{N}$, corresponding to the binding energy of the bound magnon pairs \cite{Nawa_2013}.

The theoretical prediction for the existence of a long-range ordered spin-nematic phase in \lcvo~at high fields, stabilized by the presence of a sizable inter-chain coupling, was made by Zhitomirsky and Tsunetsugu \cite{Zhitomirsky_2010}. Below the saturation field, the conditions for the appearance of a spin-nematic state are fulfilled: gapped magnon excitations and an attractive interaction between them due to the ferromagnetic exchange interaction $J_{1}$. Under these circumstances, the energy of the two-magnon bound state lies below the one of the single-magnon state, thereby stabilizing the spin-nematic phase.

Although \lcvo~ has been extensively studied, the existence of the spin-nematic phase is still under debate. A principal experimental obstacle is the very high saturation fields $H_{\texttt{sat}}$, around 45~T for  $H \parallel c$ and 52~T for  $H \parallel b$ and  $H \parallel a$. Therefore, experimental studies require the highest available DC fields or even pulsed magnetic fields. Following the first prediction for the existence of the spin-nematic state in \lcvo~\cite{Zhitomirsky_2010}, pulsed-field magnetization measurements \cite{Svistov_2011} indicated a phase occurring 4-5~T below the saturation field $H_{\texttt{sat}}=45$~T (52~T) for  $H \parallel c$ ($H \parallel b$ and  $H \parallel a$), which was attributed to a spin-nematic state. However, DC-field nuclear magnetic resonance (NMR) studies up to  45~T, on a sample from the same batch, question this interpretation and show evidence for an inhomogeneous spin state induced by non-magnetic defects, occurring in most of the field range that was previously attributed to the spin-nematic state by the magnetization data \cite{Buttgen_2014}. Whether the elimination of defects could stabilize the spin-nematic phase remains an interesting unsolved problem.

Motivated by this challenging open question, we performed NMR experiments on a high-quality \lcvo~single crystal in \textit{pulsed} magnetic fields up to 55~T providing access to the saturation fields not only for $H \parallel c$, but also for the perpendicular, $H \parallel b$, orientation. Our measurements of the NMR line position and width allow for a very precise determination of the field dependence of the local distribution of the magnetization near $H_{\texttt{sat}}$. The spin-nematic state is a homogeneous, field-dependent, longitudinal spin state without any transverse dipolar order, thus corresponding to a field-dependent NMR line position without any change of its width with respect to the saturated phase. Our NMR results in \lcvo, together with the bulk magnetization measurements \cite{SM}, perfectly match these predictions; they thus provide the first microscopic experimental evidence for the existence of a spin-nematic phase.

\begin{figure}[t!]
\centering
\includegraphics[width=0.9\columnwidth,clip]{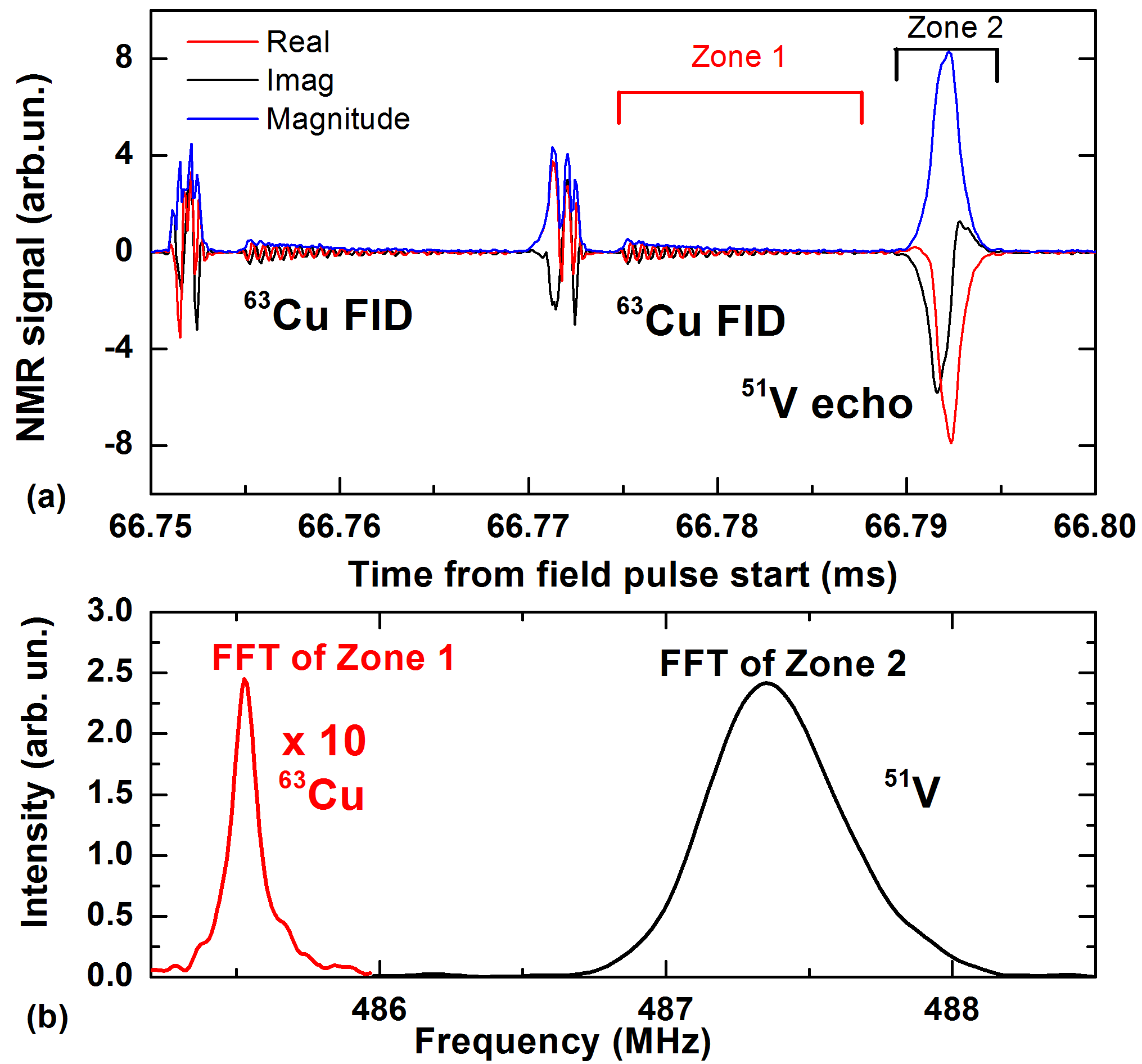}
\caption{(a): Simultaneous NMR time record of the $^{63}$Cu-metal FID and $^{51}$V spin-echo of \lcvo~ for $H \parallel c$ at $\mu_{0}H=42.91$~T and a resonance frequency of 487.2~MHz. The two $^{63}$Cu FID signals are preceded by the strong transients from radio-frequency pulses saturating the receiver. (b): Fourier transforms of the NMR time record, separately applied to Zone 1 and Zone 2 to provide the NMR spectra of $^{63}$Cu and $^{51}$V, used for field reference and the determination of the local field in ~\lcvo, respectively.}
\label{fig:1}
\end{figure}

\begin{figure*}[ht!]
\centering
\includegraphics[height=0.9\columnwidth,clip]{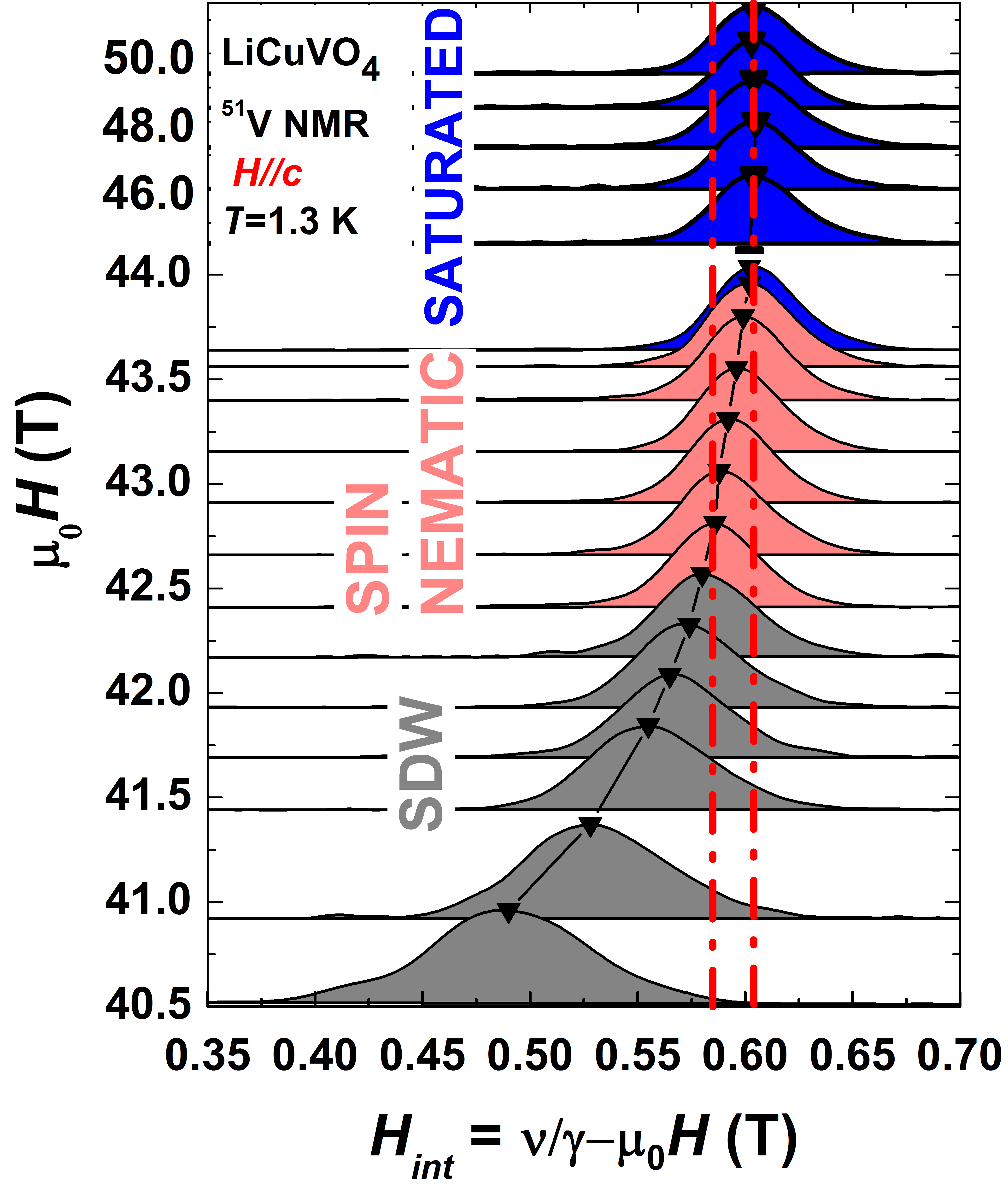}
\includegraphics[height=0.9\columnwidth,clip]{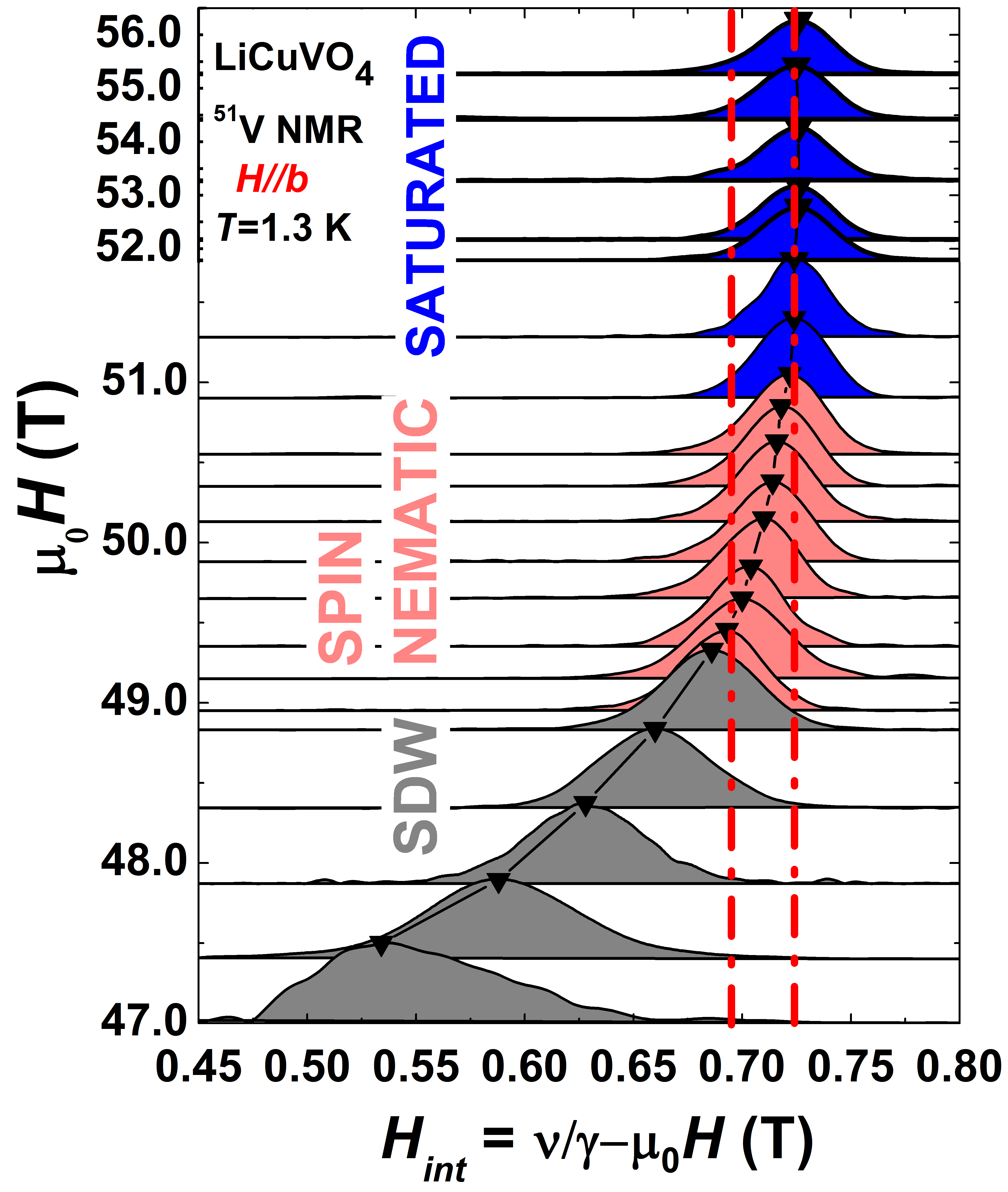}
\caption{Field dependence of the  $^{51}$V NMR spectra in \lcvo~ for $H \parallel c$ (left) and $H \parallel b$ (right) at $T = 1.3$~K, normalized to their peak intensity. The black triangles mark the peak of each NMR line, demonstrating their shift towards the saturated state. Three different regions are marked: saturated (blue), spin density wave (grey) and spin nematic (light red). The dash-dotted red lines denote the region where $H_{\rm{int}}$ becomes field dependent, but maintains the same distribution as in the saturated state.}
\label{fig:2}
\end{figure*}

NMR measurements were performed at $T$ = 1.3~K on a 2.5$\times$2$\times$0.3\,mm$^3$ \lcvo~single crystal. The studied sample was taken from the same batch that was used for the neutron-scattering experiments \cite{Mourigal_2012}. As in the previous high-field study \cite{Buttgen_2014}, the $^{51}$V (nuclear spin $I=7/2$) nuclei at the non-magnetic sites were used as a probe of the local magnetic properties of the Cu$^{2+}$ moments. The method and the experimental set-up for the NMR measurements in pulsed magnetic field have been discussed elsewhere \cite{Meier_2012,Orlova_2016,Stork_2013}. Transient pulsed magnetic fields up to 55~T with 70~ms rise time, generated by a new homogeneous pulsed-field magnet \cite{Orlova_2016}, were applied parallel to the $c$ and $b$ axes of the crystal. At the desired value of the pulsed external field, $H$, the NMR signal was recorded during a short time slot, with echo-pulse sequences consisting of two  0.5~$\mu$s excitation pulses separated by 15~$\mu$s. The very short duration of the NMR pulses ensures a spectral excitation bandwidth of 1.2~MHz, which is sufficient to record the entire $^{51}$V spectrum by a single data acquisition.

Unlike in usual experiments, NMR in pulsed-field magnets requires a calibration of the \textit{instantaneous} value of the (time-dependent) external field $H$ within the time slot of the NMR experiment. Therefore, an internal NMR reference signal has to be simultaneously recorded. Here, we use the $^{63}$Cu NMR signal from copper-metal powder placed in the same radio-frequency NMR coil together with the sample. Because of the close proximity of the gyromagnetic ratios for the $^{51}$V ($^{51}\gamma=11.199$ MHz/T) and $^{63}$Cu ($^{63}\gamma=11.285$ MHz/T) nuclei, each single NMR acquisition contains both signals (Fig.~\ref{fig:1}a): the spin-echo signal of $^{51}$V and the free-induction decay (FID) signal of the $^{63}$Cu metal. Fourier transforms of the corresponding time zones provide the $^{63}$Cu and $^{51}$V NMR spectra shown in Fig.~\ref{fig:1}b. The frequency of the $^{51}$V NMR line position  $\nu(^{51}$V) in \lcvo ~is related to the total local magnetic field $\mu_{0}(H+H_{\rm{int}})=\nu(^{51}$V$)/^{51}\gamma$, where $H_{\rm{int}}$ is the internal local field generated by the transferred hyperfine coupling from the neighboring Cu$^{2+}$ moments \cite{Nawa_2013}. $H_{\rm{int}}$ directly measures the local magnetization $M$, and is thus extracted using $H_{\rm{int}}=\nu(^{51}$V$)/^{51}\gamma-\mu_{0}H$, where $H$ is obtained from $^{63}$Cu spectral line position, $\mu_{0}H=\nu(^{63}$Cu$)/[^{63}\gamma(1$\,$+$\,$^{63}K)]$, $^{63}K$~=~0.238\% being the Knight shift of metallic copper. The large signal-to-noise ratio of the $^{51}$V and $^{63}$Cu signals and the small linewidth of the $^{63}$Cu NMR line (60~kHz) confer to these \textit{pulsed-field} measurements of $H_{\rm{int}}$ a very high precision, equivalent to what is detected in conventional constant-field NMR magnets.

Figure~\ref{fig:2} shows the field dependence of the $^{51}$V spectra for two crystal orientations, $H \parallel c$ and $H \parallel b$, at $T = 1.3$~K. The spectra taken at the same field value on the rising and falling side of the field pulse are found to be identical, and furthermore independent on the pulse-rise time (varied by 20\%). This excludes the presence of non-equilibrium phases due to the transient magnetic field. The overall behavior of the NMR spectra is similar for both orientations, where we distinguish three regions showing different behavior:
\\
i) At high fields the spectra are field \textit{independent} and consist of narrow and symmetric lines. Such behavior is observed above 43.55~T for $H \parallel c$ and above 50.55~T for $H \parallel b$. This is characteristic for a saturated homogeneous magnetic phase.
\\
ii) Below 42.41~T for $H \parallel c$ and 48.95~T for $H \parallel b$ appears a strong line-broadening; both linewidth and line position are field dependent, which is consistent with the previously identified SDW state \cite{B_ttgen_2010,B_ttgen_2012,Nawa_2013}. This phase is characterized by a modulated spin polarization, where the moments are collinear with the external field. This leads to a large distribution of the local magnetic field, causing the observed line broadening. With increasing magnetic field the width of the line and its asymmetry are decreasing due to the collapse of the SDW.
\\
iii) In the field ranges between 42.41 and 43.55~T for $H \parallel c$ and between 48.95 and 50.55~T for $H \parallel b$, the line positions change with $H$ as in the SDW phase, but their widths remain unchanged relative to those of the saturated phase. This behavior is new and has not been observed in previous NMR studies \cite{Buttgen_2014}. It clearly corresponds to the formation of a \textit{homogeneous} magnetic state as expected for a spin-nematic state.

In general, spins (\textbf{\textit{S}}) bear both a longitudinal $M_z = g_z \mu_B\langle S_z\rangle$ and a transverse component of the magnetization $M_\perp= g_\perp\mu_B\langle S_\perp\rangle$, where $g_{z}$ and $g_{\perp}$ are the corresponding components of the $g$-tensor. Therefore, both longitudinal $\langle S_{z}\rangle$ and transversal $\langle S_{\perp}\rangle$ components of the spin define $H_{\rm{int}}$ through the corresponding components of the hyperfine coupling tensor, $A_{zz}$, $A_{z\perp}$, and the g-tensor.
\bea
H_{\rm int}=\sum_{\textit{i}}[A_{zz}^{\rm (\textit{i})} g_{z}\mu_B \langle S_{z}^{\rm (\textit{i})}\rangle\ +A_{z\perp}^{\rm (\textit{i})} g_{\perp}\mu_B\langle S_{\perp}^{\rm (\textit{i})}\rangle],
\label{eq:Hint}
\eea
where $i$ runs over the 4 spins neighboring the V site (located on the top of the rectangular pyramid formed by four spin-1/2 Cu$^{2+}$ ions). For zero average $\langle S_{\perp}\rangle$ and homogeneous $\langle S_{z}\rangle$, Eq.~\eqref{eq:Hint} reduces to
\bea
H_{\rm int}=4A_{zz}g_{z}\mu_B \langle S_{z}^{\rm }\rangle,
\label{eq:Hint_av}
\eea
where  $A_{zz}$ is the value per spin unit. For further analysis we use i) the peak positions of the $^{51}$V spectra from Fig.~\ref{fig:2} to represent the local average field values $\langle H_{\rm int}\rangle$ and ii) the full width at half maximum of the line, $\Delta H_{\rm{int}}$, to characterize the spread of the local-field distribution. From Eq.~\eqref{eq:Hint_av}, we first determine the saturated spin values for $H>H_{sat}$, $\langle  S_{z}^{\rm sat} \rangle = \langle H_{\rm{int}}^{\rm sat}\rangle/(4A_{zz} g_{z}\mu_B)$. Using the observed saturation fields  $\langle H_{int,c}^{sat} \rangle = 0.605$~T and $\langle H_{int,b}^{sat} \rangle = 0.725$~T, the $A_{zz}$ values ($A_{cc}$ = 0.129 T/$\mu_B$, $A_{bb}$ = 0.166 T/$\mu_B$) determined independently in the paramagnetic state \cite{Nawa_2013}, and $g_{z}$ ($g_{c}$ = 2.313, $g_{b}$ = 2.095) from \cite{Krug_von_Nidda_2002}, we find $\langle  S_{z}^{\rm sat,c} \rangle = 0.505$ and $\langle  S_{z}^{\rm sat,b} \rangle = 0.521$ for $H \parallel c$ and $H \parallel b$, respectively. These values are close to 0.5 expected for the saturation value of a copper spin.

\begin{figure}[t!]
\centering
\includegraphics[width=0.9\columnwidth,clip]{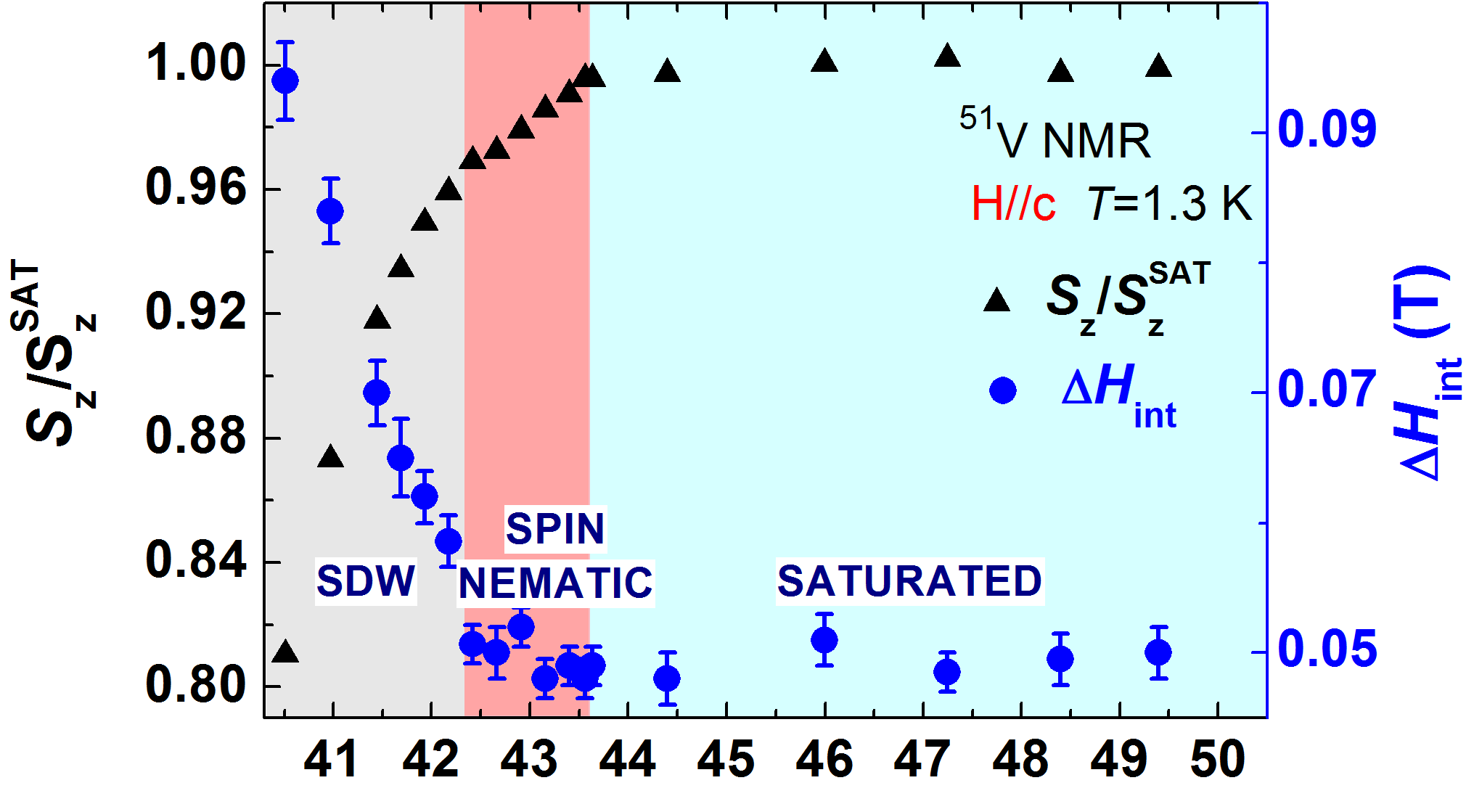}
\includegraphics[width=0.9\columnwidth,clip]{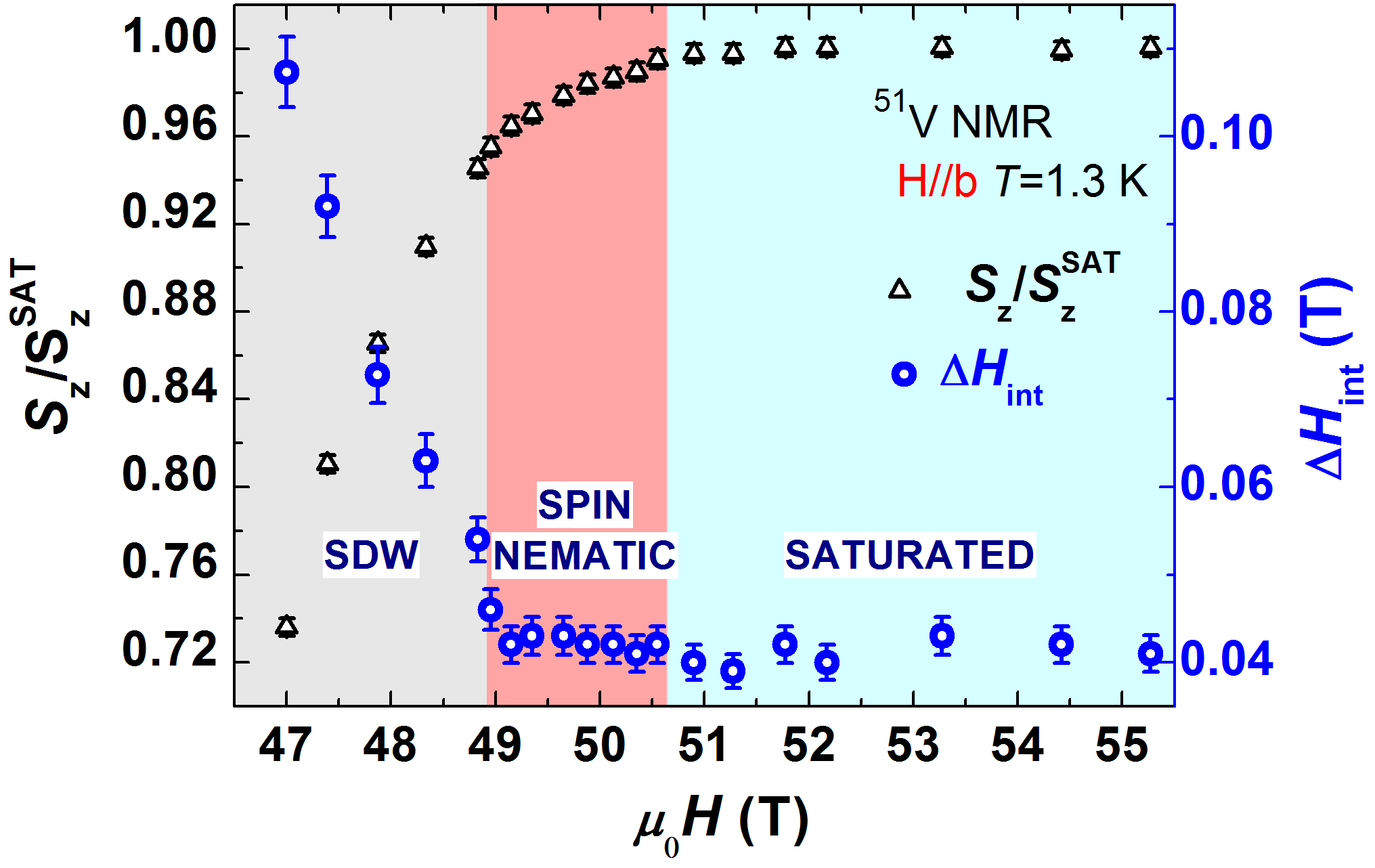}
\caption{Field dependence of the normalized spin polarization  $S_{z}/S_{z}^{\rm sat}$ (solid and open black triangles) and distribution widths of the internal magnetic field  $\Delta H_{\rm{int}}$ (solid and open blue circles) obtained from the $^{51}$V NMR spectra in \lcvo~ shown in Fig. 2, for $H \parallel c$ (top) and $H \parallel b$ (bottom). Three characteristic regions are marked by background colors: the SDW (grey), spin-nematic (red) and saturated (blue) phase.}
\label{fig:3}
\end{figure}

We then plot  in Fig.~\ref{fig:3} the field dependence of the local spin polarization $S_{z}$ normalized by $S_{z}^{\rm sat}$ and the spread of the local-field distribution $\Delta H_{\rm{int}}$.
These data summarize the behavior of the NMR spectra and allow us to obtain more quantitative information, in particular for the intermediate, nematic phase, where $S_{z}$ changes with field while $\Delta H_{\rm{int}}$ remains constant, keeping the same value as in the saturated phase. The total change of the spin polarization is there about 4\%, within a field range of 1.2~T for $H \parallel c$ and 1.6~T for $H \parallel b$.
As $\Delta H_{\rm{int}}$ is sensitive to both $\langle S_{z}\rangle$ and $\langle S_{\perp}\rangle$ spin components, the absence of the line broadening in this field range excludes inhomogeneous transverse magnetic order, such as a conical state, and rules out any formation of inhomogeneous magnetization (distribution of magnetic moments). The coupling to $\langle  S_{\perp} \rangle$ is very strong in \lcvo, $A_{z\perp}$ = 0.12~T/$\mu_B$ \cite{Nawa_2013}, and a transverse order would thus generate a distribution or splitting of local fields and, thereby, increase $\Delta H_{\rm{int}}$ relative to the saturated phase, which is not observed here. Therefore, our results clearly show that the underlying state is homogeneous and thus preserves translation symmetry. Since these properties are characteristic for a spin-nematic state, they provide an experimental proof for its existence in~\lcvo~at high magnetic fields.

Our results fundamentally differ from previous NMR work of B{\"u}ttgen et al. \cite{Buttgen_2014}. First, below saturation they observe an inconsistency between the bulk magnetization and the NMR data. They conclude that the majority of the sample is already in the saturated phase as monitored by NMR even though the magnetization still exhibits a linear slope due to the presence of defects. This is in contrast to our result, where the bulk magnetization measured on the same sample batch approximately \emph{coincides} with the local magnetization measured by NMR, and also shows the signature of the nematic phase \cite{SM}. Second, they attribute a possible spin-nematic phase to a narrow field range between 40.5~T and 41.4~T, where the (local) magnetization exhibits a very steep slope and the NMR spectra present a continuous \emph{change} of $\Delta H_{\rm{int}}$ as a function of $H$, which signifies a developing inhomogeneous nonuniform magnetization that should be \emph{absent} in a spin-nematic phase. Our results, however, fulfill \emph{both} criteria for a spin-nematic state: a linear local $M_{z}(H)$ dependence and a \emph{constant} $\Delta H_{\rm{int}}$. The two different NMR observations might be related to different defect concentrations in the two samples, and further experimental studies appear to be necessary to clarify this point. In addition, we consistently observe the same NMR signature of the spin-nematic phase for \emph{two} crystal orientations.

A careful examination of Fig.~\ref{fig:3} reveals that there is a subtle difference in the field dependence of $S_{z}(H)$ in the spin-nematic phase for the two crystal orientations: $S_{z}$ changes linearly for $H \parallel c$, with clear kinks at the saturation field and at the transition to the SDW phase, while  $S_{z}(H)$ shows a more smooth field dependence for $H \parallel b$. This is probably related to an anisotropy effect: \lcvo ~has an easy-plane anisotropy ($ab$ plane), preserving all symmetry properties only in the $H \parallel c$ direction. Indeed, from the ESR data \cite{Krug_von_Nidda_2002} the anisotropy has been attributed to the anisotropy of the exchange coupling $J$, estimated to be nearly axially symmetric, with the $J_{cc}$ value reduced by $\approx$6\%. Within a purely 1D model \emph{without} frustration, a transverse anisotropy of $J$ induces even at zero temperature a small finite depolarization at the critical field $H_\textrm{c}$, well described by the mean-field approximation \cite{Hagemans2005}. We find $\langle S_{z}(H_\textrm{c})\rangle = \arccos(\delta)/(\pi \sqrt{1-\delta^2})$, where $\delta \approx |J_{xx} - J_{yy}|/(J_{xx} + J_{yy} + J_{zz})$. Applied to \lcvo, the depolarization would be \emph{zero} for $H \parallel c$ and 1.3\% for $H \parallel b$, meaning that it may somewhat influence the magnetization only for this latter orientation. While this is consistent with our NMR data, a complete theoretical description that includes both frustration and anisotropy remains to be done.

In summary, using $^{51}$V NMR, we microscopically characterized the high-field properties of \lcvo~for two crystal orientations, $H \parallel c$ and $H \parallel b$. Just below the full saturation, the $^{51}$V NMR spectra evidence a field range where the \textit{homogeneous} local magnetization is increasing with field. We argue that such behavior corresponds to the predicted spin-nematic phase, a state partially breaking spin-rotation symmetry around the magnetic-field axis without generating any transverse dipolar magnetic order. The experimentally observed field dependences of $S_{z}$ and their difference for $H \parallel c$ and $H \parallel b$, probably reflecting the easy-plane anisotropy in \lcvo, provide a qualitative and quantitative basis for further theoretical investigation of this material, and should help in distinguishing different models for the spin-nematic phase.

\begin{acknowledgments} We thank C. Berthier and M. Zhitomirsky for stimulating discussions. Nicolas Bruyant, Marc Nardone, Lo\"ic Drigo, J\'erome B\'eard are thanked for technical assistance at the LNCMI-Toulouse. We acknowledge support of the Deutsche Forschungsgemeinschaft (DFG) through SFB~1143.
\end{acknowledgments}

\clearpage
\renewcommand{\thefigure}{S\arabic{figure}}
\setcounter{figure}{0}
\renewcommand{\theequation}{S\arabic{equation}}
\setcounter{equation}{0}

\begin{center}
\textbf{\large SUPPLEMENTAL MATERIAL \\} \vspace{10pt}
\textbf{to: "Nuclear magnetic resonance signature of the spin-nematic phase in \lcvo~at high magnetic fields" by A.~Orlova,$^*$ E.~L.~Green,$^\dag$ J.~M.~Law,$^\ddag$ D.~I.~Gorbunov, G.~Chanda, S.~Kr\"amer, \mbox{M. Horvati\'c}, R.~K.~Kremer, J.~Wosnitza, G.~L.~J.~A.~Rikken}
\end{center}

\begin{figure}[b!]
\centering
\includegraphics[width=0.85\columnwidth,clip]{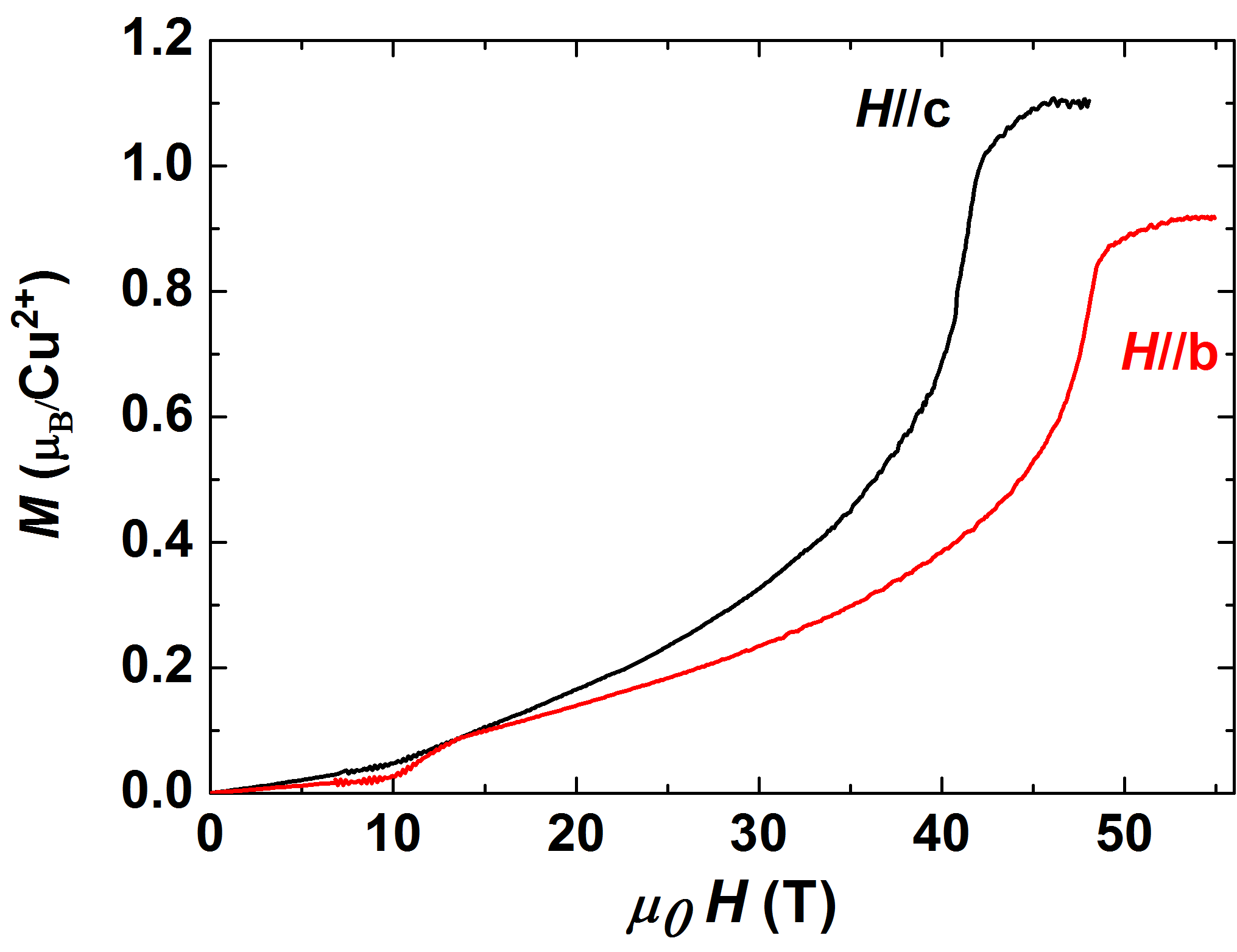}
\caption{Magnetization $M(H)$ of \lcvo~ measured for $H \parallel c$ (black) and $H \parallel b$ (red) at 1.5~K.}
\label{Sfig:4}
\end{figure}

High-field magnetization was measured at 1.5 K in pulsed magnetic fields up to 55~T  (rise time~7~ms, total pulse duration 25-35~ms) at the Dresden High Magnetic Field Laboratory, Germany. The field was applied along the $b$ and $c$ axes of a \lcvo~, a single crystal from the same sample batch that was used for the NMR experiments. The magnetization was measured by the induction method using a coaxial pick-up coil system. A detailed description of the setup is given in \cite{Skourski_2011}.
Fig.~\ref{Sfig:4} shows the the magnetization of \lcvo~ measured at 1.5~K for $H \parallel c$ and $H \parallel b$. Within the measurement precision our data agree to those previously published by Svistov et al. \cite{SvistovS_2011}.

\begin{figure}[t!]
\centering
\includegraphics[width=1\columnwidth,clip]{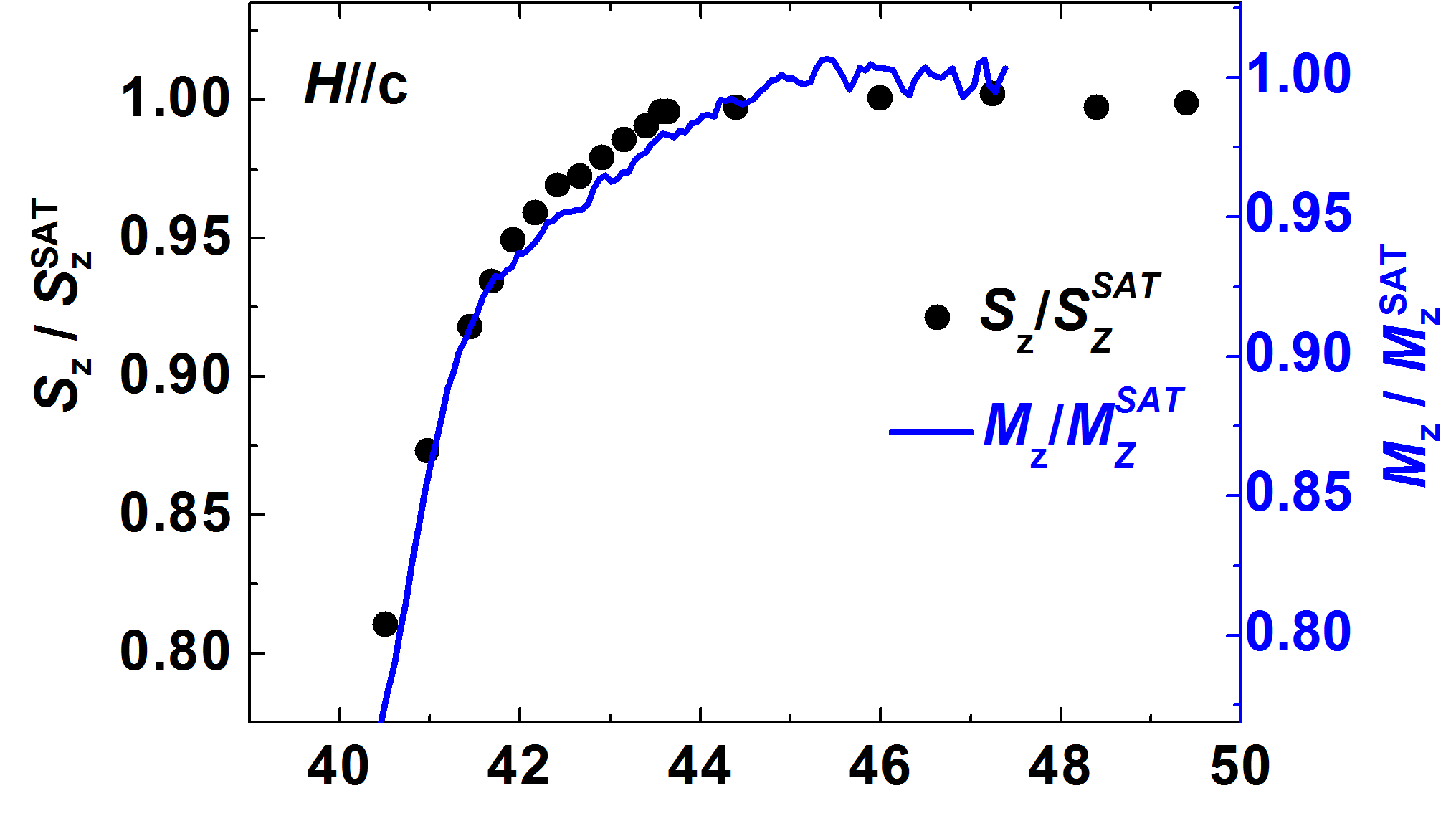}
\includegraphics[width=1\columnwidth,clip]{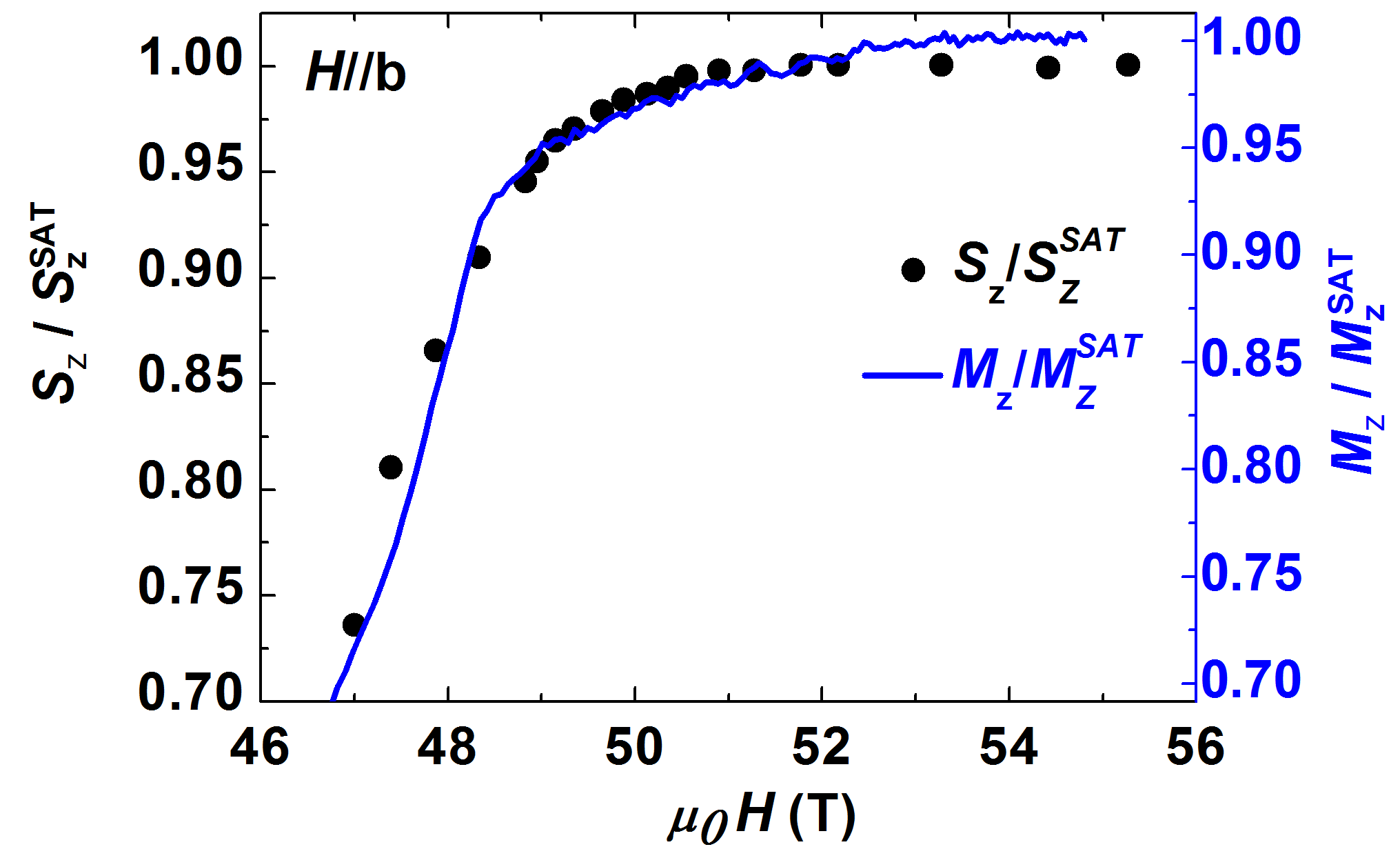}
\caption{Magnetic field dependence of the normalized spin polarization  $S_{z}/S_{z}^{\rm sat}$ determined by NMR and the normalized bulk magnetization $M_{z}/M_{z}^{\rm sat}$ (solid lines) in \lcvo~ for $H \parallel c$ (top) and $H \parallel b$ (bottom).}
\label{Sfig:5}
\end{figure}

Fig.~\ref{Sfig:5} displays the field dependence of the normalized spin polarization,  $S_{z}/S_{z}^{\rm sat}$, determined by NMR (data taken from Fig.3 of the main manuscript) together with $M_{z}/M_{z}^{\rm sat}$. The field axis of the $M(H)$ data are shifted by 0.6~T for $H \parallel c$ and by 0.15~T for $H \parallel b$. These corrections of the field scale are within the error bars of the magnetic-field determination during the magnetization experiment. The two vertical axes are scaled by 1~\%.
The good coincidence between the NMR data and $M(H)$ demonstrates that the local magnetization measured by NMR and the bulk magnetization measured by the induction method are coinciding for our \lcvo~ crystal. This fact further evidences the existence of a spin-nematic state.

\end{document}